\magnification = 1030
\baselineskip16.5pt
\abovedisplayskip 6pt plus2pt minus5pt
\belowdisplayskip 6pt plus2pt minus5pt
\vsize=9.5 true in \raggedbottom
\hsize=5.7 in
\voffset= -0.1 true in
\font\bmit=cmmib10 \textfont9=\bmit \def\bmit{\fam9 }
\font\bx=cmbsy10 \textfont10=\bx \def\bx{\fam10 }

\mathchardef\alpha="710B
\mathchardef\pi="7119
\mathchardef\sigma="711B
\mathchardef\mu="7116
\mathchardef\nabla="7272
\parskip=8pt plus3pt
\interlinepenalty=1500
\def\boxit#1{\vbox{\hrule\hbox{\vrule\kern1pc
   \vbox{\kern1pc#1\kern1pc}\kern1pc\vrule}\hrule}}
\def\sqr#1#2{{\vcenter{\vbox{\hrule height.#2pt
   \hbox{\vrule width.#2pt height#1pt  \kern#1pt
      \vrule width.#2pt}
     \hrule height.#2pt }}}}
\def\square{\mathchoice\sqr64\sqr64\sqr{4.2}3\sqr33}

\leftline{\bf Energy-momentum operators with eigenfunctions localized
along an axis}

\rightline{Shaun N Mosley,\footnote{${}^*$} {E-mail:
shaun.mosley@ntlworld.com \hfil\break
 correspondence: Sunnybank, Albert Rd, Nottingham NG3 4JD, UK}
Alumni, University of Nottingham, UK }

\vskip 0.6in

\noindent {\bf Abstract } \quad
The momentum operator
$ {\bf p} = - \, i \, {\bx \nabla} \, $ has radial component
$ {\bf \tilde p} \equiv - \, i \, {\bf \hat{r}} ( {1 \over r} \,
 \partial_r \, r ) \, .$ We show that $ {\bf \tilde p} \, $ is the space part
of a 4-vector operator, the zero component of which is a positive operator.
Their eigenfunctions are localized along an axis through the origin.
The solutions of the evolution equation
$ i \, \partial_t \, \psi = {\tilde p^0} \, \psi \, $ are waves along the
propagation axis. Lorentz transformations of these waves yield
the aberration and Doppler shift.
We briefly consider spin-half and spin-one representations.
\vskip 0.1 in

\beginsection I. Introduction

Our everyday experience of photons is that they are localized
along their propagation axis, whereas in the usual formalism a photon of
definite momentum (an eigenvalue of $ {\bf p} \, )$ is a plane wave spread
over all space. The operator $ {\bf p} = - \, i \, {\bx \nabla} \, $
can be split into a radial component $ {\bf \tilde p} \, $
and an angular (transverse) component due to the identity
$$ \eqalign{
- \, i \, {\bx \nabla}
&= - \, i \, {\bf \hat{r}} ( {1 \over r} \,
 \partial_r \, r ) \,
- \, {1 \over 2} \, {1 \over r} \, ( {\bf \hat{r}} \times {\bf L} \,
- {\bf L} \times {\bf \hat{r}} )
\equiv {\bf \tilde p} \, + \, {\bf \check p} \, \, \cr
  }  \eqno (1) $$
with $ {\bf \hat{r}} \equiv {\bf r} / r \, $ and
$ {\bf L} \, $ the angular momentum operator.
We note that
$$ \eqalign{
&{\bf \tilde p} \, \psi ( {\bf r} ) \,
\equiv - \, i \, {{\bf \hat{r}} \over r} \,
 \partial_r \, \big( r  \, \psi ( {\bf r} ) \big) \, \cr
  }  \eqno (2) $$
is undefined at the origin. In the appendix we show that continuity
in $ r\, \psi ( {\bf r} ) \, $ on opposite sides of the origin, i.e. that
$$ r \, \psi ( \epsilon {\bf r} ) -  r \, \psi ( - \epsilon {\bf r} )
\rightarrow 0 \,
\qquad \qquad \hbox{as } \qquad \epsilon \rightarrow 0 \, \eqno (3) $$
is a sufficient condition for
$ {\bf \tilde p} \, $ to be a symmetric operator with respect to the usual
inner product space
$$ \eqalignno{
&\Big\langle \psi_1 ( {\bf r} ) \, | \,
\psi_2 ( {\bf r} ) \, \Big\rangle
\equiv \int  d^3 {\bf r} \; \psi_1^* ( {\bf r} ) \,
\psi_2 ( {\bf r} )
= \int_\Omega d\Omega \int_0^\infty dr \; r^2 \,
\psi_1^* ( r,\theta , \phi )  \,
\psi_2 ( r,\theta , \phi )  \,  . & (4) \cr
  }  $$

The eigenfunction of $ {\bf \tilde p} \, $ with eigenvalue $ {\bf k} \, $ is
$$ \eqalignno{
u_{\bf k} &= \, {1 \over \sqrt{2 \pi} \,k r } \, [ \, \delta ( {\bf \hat{r}} ,
{\bf \hat{k}} ) \, \exp ( i k r ) \, + \, \delta ( - {\bf \hat{r}} ,
{\bf \hat{k}} ) \, \exp ( - i k r ) \, ] \,         & (5a)  \cr
 &= \, {1 \over \sqrt{2 \pi} \,k r } \,
[ \, \delta ( {\bf \hat{r}} , {\bf \hat{k}} ) \, + \, \delta ( - {\bf \hat{r}} ,
{\bf \hat{k}} ) \, ] \,
\exp ( i {\bf k} \cdot {\bf r} ) \,         & (5b)  \cr
\noalign{\noindent where $ {\bf \hat{k}} \equiv {\bf k} / |{\bf k}| \, ,$
$ k \equiv |{\bf k}| \, $ and
$ \delta ( {\bf \hat{r}} , {\bf \hat{k}} ) \, $
is the delta function on the unit sphere (for this notation see the appendix).
The equivalence between (5a) and (5b) may be seen by writing
$ \delta ( {\bf \hat{r}} , {\bf \hat{k}} ) \, \exp ( i k r )
=  \delta ( {\bf \hat{r}} , {\bf \hat{k}} ) \, \exp ( i k ( {\bf \hat{r}}
\cdot {\bf \hat{r}} ) r )
=  \delta ( {\bf \hat{r}} , {\bf \hat{k}} ) \, \exp ( i k ( {\bf \hat{k}}
\cdot {\bf \hat{r}} ) r )  \,
=  \delta ( {\bf \hat{r}} , {\bf \hat{k}} ) \, \exp ( i {\bf k}
\cdot {\bf r} )  \, ,$ and similarly for
$  \delta ( - {\bf \hat{r}} , {\bf \hat{k}} ) \, \exp ( - i k r ) \, .$
Then from (5a) }
{\bf p}  \, u_{\bf k}
&= \, {1 \over \sqrt{2 \pi} \,k r } \, [
\, {\bf \hat{r}} \, k \,\delta ( {\bf \hat{r}} ,
{\bf \hat{k}} ) \, \exp ( i k r ) \, - \, {\bf \hat{r}} \, k \,
\delta ( - {\bf \hat{r}} , {\bf \hat{k}} ) \, \exp ( - i k r ) \, ] \,  \cr
&= \, {1 \over \sqrt{2 \pi} \,k r } \, [
\, {\bf \hat{k}} \, k \,\delta ( {\bf \hat{r}} ,
{\bf \hat{k}} ) \, \exp ( i k r ) \, + \, {\bf \hat{k}} \, k \,
\delta ( - {\bf \hat{r}} , {\bf \hat{k}} ) \, \exp ( - i k r ) \, ] \,
= {\bf k} \, u_{\bf k} \, .               & (6)  \cr
  } $$
The state $ u_{\bf k} \, $ is localized along the entire
$ {\bf k} \, $ axis $ {\bf r} =  \lambda \, {\bf \hat{k}} \;
( - \infty < \lambda < \infty ) \, ,$ and the density
$ ( u^*_{\bf k} u_{\bf k} ) \, $ is evenly distributed along the
$ {\bf k} \, $ axis as
$$ \int_{r = a}^b ( u^*_{\bf k} u_{\bf k} ) \, d^3{\bf r}
= (b - a ) / \pi k^2 \, .$$
Note that
$$ {\cal P} u_{\bf k} = u_{\bf k}^*
= u_{ \{ {\bf k} \rightarrow - {\bf k} \} } \, $$
where $ {\cal P} \, $ is the parity operator.

The $ u_{\bf k} \, $ satisfy the orthogonality and completeness relations
$$ \eqalignno{
&\int d^3 {\bf r} \; u^*_{\bf k} ( {\bf r} ) \, u_{\bf k'}  ( {\bf r} )
= \, \delta ({\bf k} - {\bf k'} ) \, , \qquad \qquad
\int d^3 {\bf k} \; u^*_{\bf k} ( {\bf r} ) \, u_{\bf k} ( {\bf r'} )
= \, \delta ({\bf r} - {\bf r'} ) \, & (7)  \cr
  }  $$
which are verified in the appendix. As the Lorentz invariant measure
is $ d^3 {\bf k} / k \, $ rather than $ d^3 {\bf k} \, ,$ this suggests
defining the inner product spaces
$$ \eqalignno{
&\Big\langle \phi_1 ( {\bf k} ) \, | \,
\phi_2 ( {\bf k} ) \, \Big\rangle_{1/k}
\equiv \int { d^3 {\bf k} \over k } \; \phi_1^* ( {\bf k} ) \,
\phi_2 ( {\bf k} ) \, , & (8) \cr
&\Big\langle \psi_1 ( {\bf r} ) \, | \,
\psi_2 ( {\bf r} ) \, \Big\rangle_{1/r}
\equiv \int { d^3 {\bf r} \over r } \; \psi_1^* ( {\bf r} ) \,
\psi_2 ( {\bf r} ) \, , & (9) \cr
  }  $$
and the basis state
$$ \eqalignno{
w_{\bf k} = \sqrt{ k r } \, u_{\bf k} \,
&=\, {1 \over \sqrt{ 2 \pi k r } } \, [ \, \delta ( {\bf \hat{r}} ,
{\bf \hat{k}} ) \, \exp ( i k r ) \, + \, \delta ( - {\bf \hat{r}} ,
{\bf \hat{k}} ) \, \exp ( - i k r ) \, ] \,     & (10a)  \cr
&=\, {1 \over \sqrt{ 2 \pi k r } } \, [ \, \delta ( {\bf \hat{r}} ,
{\bf \hat{k}} ) \, + \, \delta ( - {\bf \hat{r}} ,
{\bf \hat{k}} ) \, ] \, \exp ( i \, {\bf k} \cdot {\bf r} ) \, ,    & (10b)  \cr
   }  $$
which is an eigenfunction of the operator
$$ {\bf \overline p} \equiv - \, i \, {\bf \hat{r}} ( {1 \over \sqrt{ r}} \,
 \partial_r \, \sqrt{ r} ) \, = - \, i \, {\bf \hat{r}}
( \partial_r + {1 \over 2 r} ) \,  \eqno (11) $$
which is symmetric with respect to the $ 1 / r \, $ inner product space (9).
From now on we will be concerned with the operator $ {\bf \overline p} \, $
and its eigenfunctions $ w_{\bf k} \, $ (rather than with
$ {\bf \tilde p} , \, u_{\bf k} \, )$ because $ w_{\bf k} \, $ transforms
as a scalar under Lorentz transformations (see below Sec. 3).
From (7)
$$ \eqalignno{
&\big\langle w_{\bf k} ( {\bf r} ) \,  | \, w_{\bf k'}  ( {\bf r} ) \, \big\rangle_{1/r}
= \, k \, \delta ({\bf k} - {\bf k'} ) \, , \qquad \qquad
\big\langle w_{\bf k} ( {\bf r} ) \,  | \,
w_{\bf k}  ( {\bf r'} ) \, \big\rangle_{1/k}
= \, r \, \delta ({\bf r} - {\bf r'} ) \, . & (12) \cr
  }  $$

Given a $ \psi ({\bf r}) \, $ which is a superposition of various
$ w_{\bf k} \, $
$$ \psi ({\bf r}) = \int { d^3 {\bf k} \over k } \;
\phi ({\bf k}) \, w_{\bf k}
\equiv \Big\langle  w_{\bf k}^* \, | \, \phi ( {\bf k} ) \, \Big\rangle_{1/k}   \, \eqno (13) $$
then the orthogonality relation (12) enables one to project out
the distribution $ \phi ({\bf k}) \, $ as
$$ \Big\langle  w_{\bf k} \, | \, \psi ( {\bf r} ) \, \Big\rangle_{1/r}
=  \Big\langle  w_{\bf k} \, | \, \int { d^3 {\bf k'} \over k' } \;
\phi ({\bf k'}) \, w_{\bf k'}  \, \Big\rangle_{1/r}
= \phi ({\bf k}) \, . $$
Expanding out the quantity $ \Big\langle  w_{\bf k} \, | \,
\psi ( {\bf r} ) \, \Big\rangle_{1/r} \, $ we show that this is a Fourier
transform of $ \psi ( {\bf r} ) \, $ as well as of
$ \psi ( - {\bf r} ) \, $ along the line
$ {\bf r} =  \lambda \, {\bf \hat{k}} \, ,\;
( - \infty < \lambda < \infty ) ,$ as
$$ \eqalignno{
\Big\langle  w_{\bf k} \, | \,
\psi ( {\bf r} ) \, \Big\rangle_{1/r} \, = \phi ({\bf k}) \,
&\equiv \int { d^3 {\bf r} \over r } \; \psi ( {\bf r} ) \,
 \, w_{\bf k}^* ( {\bf r} )              \cr
&= \int_0^\infty dr \, r \int_\Omega d\Omega \; \psi ( r {\bf \hat{r}} ) \,
{1 \over \sqrt{ 2 \pi k r } } \,
\big[ \, \delta ( {\bf \hat{r}} , {\bf \hat{k}} ) \, \exp ( - i k r ) \,
+ \, \delta ( - {\bf \hat{r}} , {\bf \hat{k}} ) \, \exp ( i k r ) \,
\big] \,  \cr
&= {1 \over \sqrt{ 2 \pi } } \,
\int_0^\infty dr \; \sqrt{ r \over k} \;
\big[ \,  \psi (  r {\bf \hat{k}} ) \, \exp ( - i k r ) \,
+ \, \psi ( - r {\bf \hat{k}} ) \, \exp ( i k r ) \, \big] \, . \cr
\noalign{\noindent It is more transparent here to write $ {\bf k} \, $ in
spherical
coordinates: with $ {\bf k} = (k , \theta_k , \phi_k ) \, $ then the above is }
\phi (k , \theta_k , \phi_k ) \,
&= {1 \over \sqrt{ 2 \pi } } \,
\int_0^\infty dr \; \sqrt{ r \over k} \;
\big[ \,  \psi (r , \theta_k , \phi_k )  \, \exp ( - i k r ) \,
+ \, \psi (r , \pi - \theta_k , \phi_k + \pi ) \, \exp ( i k r ) \, \big] \,
\cr
&= {1 \over 2 \sqrt{ k } } \,
\big[ ( {\cal F}_c \, - \, i \, {\cal F}_s \, ) \,
+ \, ( {\cal F}_c \, + \, i \, {\cal F}_s \, ) \, {\cal P} \, \big]
\, \sqrt{r} \, \psi (r , \theta_k , \phi_k ) \,  & (14) \cr
&= {\cal U}  \, \psi (r , \theta_k , \phi_k ) \,  \cr
  }  $$
where $ {\cal P } \, $ is the parity operator and
$ {\cal F}_c , \, {\cal F}_s \, $ are the Fourier cosine, sine transforms
defined by
$$ \eqalign{
{\cal F}_c \, f(r , \theta , \phi )
= g_c (k , \theta , \phi ) &\equiv \, \sqrt{2 \over \pi} \,
\int_0^\infty  f(r , \theta , \phi ) \, \cos (r k ) \, dr \, , \cr
{\cal F}_s \, f(r , \theta , \phi )
= g_s (k , \theta , \phi ) &\equiv \, \sqrt{2 \over \pi} \,
\int_0^\infty  f(r , \theta , \phi ) \, \sin (r k ) \, dr \, . \cr
  }    \eqno (15) $$
The operator of (14)
$$ {\cal U} \equiv {1 \over 2 \sqrt{ k } } \,
\big[ ( {\cal F}_c \, - \, i \, {\cal F}_s \, ) \,
+ \, ( {\cal F}_c \, + \, i \, {\cal F}_s \, ) \, {\cal P} \, \big]
\, \sqrt{r} \, \eqno (16) $$
defines a unitary mapping
from $ \psi ( {\bf r} ) \, $ to $ \phi ( {\bf k} ) \, $
whose inverse is
$$ \tilde{\cal U} \equiv {1 \over 2 \sqrt{ r } } \,
\big[ ( \tilde{\cal F}_c \, + \, i \, \tilde{\cal F}_s \, ) \,
+ \, ( \tilde{\cal F}_c \, - \, i \, \tilde{\cal F}_s \, ) \, {\cal P} \, \big]
\, \sqrt{k} \, \eqno (17) $$
where the $ \tilde{\cal F}_c \, \tilde{\cal F}_c \, $ are as in (15)
but with $ r , k \, $ interchanged. It is straightforward to check that
$ \tilde{\cal U} \, {\cal U} \, = 1 \, ,$ with the aid of
$ \tilde{\cal F}_c \, {\cal F}_c   = \tilde{\cal F}_s \, {\cal F}_s
= {\cal P} \, {\cal P} \, = 1 \, .$
We emphasize that
$ {\cal U} , \, \tilde{\cal U} \, $ are in a sense one dimensional transform
operators:
$ {\cal U} \, $ maps $ \psi ( {\bf r} ) \, $ on any axis through the origin
onto $ \phi ( {\bf k} ) \, $ on the same axis.

As $ {\bf \overline p } \, w_{\bf k} \, = {\bf k} \, w_{\bf k} \, ,$
and $ {\bf k} \, $ is the space part of the 4-vector $ k^\lambda
\equiv ( k , {\bf k} ) \, ,$
we look for the operator $ \overline p^0 \, $ such that
$ \overline p^0 \, w_{\bf k} = k \, w_{\bf k} \, .$
(One can see by inspection of (10a) that $ w_{\bf k} \, $ is not an
eigenfunction of the ``radial momentum operator''
$ p_r \equiv - i r^{-1/2} \partial_r r^{1/2}  \, $
which anyway is not a Hermitian operator [1]). The operator
$ \, \overline p^0 \, $ is multiplication by  $ k \, $ in momentum space,
i.e.
$$\eqalignno{
 \overline p^0 \, &= \, \tilde{\cal U} \,  k \,
{\cal U} \,          & (18)  \cr    }  $$
so that
$ \, \overline p^0 \, $ is a positive operator.
To simplify (18) we need the following identities [2]
$$ \eqalign{
\tilde{\cal F}_c {\cal F}_c \, = \, \tilde{\cal F}_s {\cal F}_s \, = \, 1
\, , \qquad
\tilde{\cal F}_s {\cal F}_c \, = \, - \, {\cal H}_e \, , \qquad
\tilde{\cal F}_c {\cal F}_s \, = \, {\cal H}_o \, \cr
  } \eqno (19)  $$
where $ {\cal H}_e , \, {\cal H}_o \, $ are the Hilbert transforms of
even, odd functions defined by
$$ \eqalign{
{\cal H}_e \, f(r , \theta , \phi ) &\equiv \, - \, { 2 r \over \pi } \,
\int_0^\infty { f(t , \theta , \phi ) \over r^2 - t^2 } \, dt \, , \qquad
{\cal H}_e \, f({\bf r}) = \, - \, { 2  \over \pi } \,
\int_0^\infty { \, f( \lambda {\bf r} ) \over 1 - \lambda^2 } \,
d\lambda \, , \cr
{\cal H}_o \, f(r , \theta , \phi ) &\equiv  \, - \,  { 2 \over \pi } \,
\int_0^\infty { t \, f(t , \theta , \phi ) \over r^2 - t^2 } \, dt \, ,
\qquad \quad
{\cal H}_o \, f({\bf r}) =  \, - \,  { 2 \over \pi } \,
\int_0^\infty { \lambda \, f( \lambda {\bf r} ) \over 1 - \lambda^2 } \,
d\lambda \, . \cr }
  \eqno (20)   $$
We will write
$$  {\cal F}_\pm \equiv {\cal F}_c \, \pm \, i \, {\cal F}_s \, $$
then from (19) there follows the further identities
$$ \eqalignno{
&\overline{\cal F}_+ \, {\cal F}_+
= \, - \, i \, ( {\cal H}_e - {\cal H}_o )  \, , \qquad \qquad
\overline{\cal F}_- \, {\cal F}_-
= \, i \, ( {\cal H}_e - {\cal H}_o )  \, , \qquad  \cr
&\overline{\cal F}_+ \, {\cal F}_-
= 2 \, - \, i \, ( {\cal H}_e + {\cal H}_o )  \, , \qquad
\overline{\cal F}_- \, {\cal F}_+
= 2 \, + \, i \, ( {\cal H}_e + {\cal H}_o )  \, , \qquad     \cr
\noalign{\noindent also }
&\overline{\cal F}_\pm \, k
= \mp \, i \, \partial_r \, \overline{\cal F}_\pm  \, , \qquad \qquad  \qquad
k \, {\cal F}_\pm \,
= \pm \, i \, {\cal F}_\pm  \, \partial_r \, . \qquad  & (21) \cr
 }       $$
Returning to $ \overline p^0 \, = \, \tilde{\cal U} \,  k \, {\cal U} \, $
we have
$$ \eqalignno{
\overline p^0 \, = \, \tilde{\cal U} \,  k \, {\cal U} \,
&=  {1 \over 4 \sqrt{r} } \,
\left( \overline{\cal F}_+ \, + \, \overline{\cal F}_-  {\cal P} \,
 \right) \, k \, \left( {\cal F}_-  \, + \, {\cal F}_+  {\cal P} \,
 \right) \, \sqrt{r} \, \cr
&= \, i \, {1 \over 4 \sqrt{r} } \, \partial_r \,
\left( - \, \overline{\cal F}_+  \,
+ \, \overline{\cal F}_- {\cal P} \, \right) \,
\left( {\cal F}_-  \, + \, {\cal F}_+  {\cal P} \,  \right) \,  \sqrt{r} \, \cr
&= \, i \, {1 \over 4 \sqrt{r} } \, \partial_r \,
\left[ ( - \, \overline{\cal F}_+ {\cal F}_-  \,
+ \, \overline{\cal F}_- {\cal F}_+  ) \,
+ \, ( \, \overline{\cal F}_- {\cal F}_-
- \, \overline{\cal F}_+ {\cal F}_+   ) \, {\cal P} \, \right]  \sqrt{r} \, \cr
&= \, - \, {1 \over 2 \sqrt{r} } \, \partial_r \,
\left[ \, ( {\cal H}_e + {\cal H}_o )  \,
+ \,  ( {\cal H}_e - {\cal H}_o )  \, {\cal P} \, \right]  \,
\sqrt{r} \,
= \, - \, {1 \over \sqrt{r} } \, \partial_r \, {\cal H}_+   \sqrt{r} \,
      & (22) \cr
   }  $$
where
$$ \eqalignno{
{\cal H}_+  f({\bf r})
&\equiv \, {1 \over 2} \, \left[ \, ( {\cal H}_e + {\cal H}_o )  \,
+ \,  ( {\cal H}_e - {\cal H}_o )  \, {\cal P} \, \right] \, f({\bf r})
  &  (23a) \cr
&= \,  { 1 \over \pi } \, \int_0^\infty
\Big( \,  {  f( \lambda {\bf r}) \over \lambda - 1 } \,
- \, {  f( - \lambda {\bf r}) \over 1 + \lambda  } \, \Big) \;
d\lambda \,   \cr
&= \,  { 1 \over \pi } \, \int_{- \infty}^\infty
 \,  {  f( \lambda {\bf r}) \over \lambda - 1 }  \;
d\lambda \, &  (23b)   \cr
     }  $$
which is the Hilbert transform of $ f ({\bf r}) \, $ along the axis
$  \lambda \, {\bf r} \, ,\;
( - \infty < \lambda < \infty ) \, .$

We now establish that
$ \overline p^0 \, w_{\bf k} \, \equiv \, - \, {1 \over \sqrt{r} } \,
\partial_r \, {\cal H}_+   \sqrt{r} \,  w_{\bf k} \,
= k \, w_{\bf k} \, .$ For simplicity we will choose the particular case
when
$ {\bf k} = ( 0 , 0 , k ) \, , \; k > 0 \, .$ Then
$$ \eqalignno{
w_{k^3} &=\, {1 \over \sqrt{ 2 \pi k r } } \,
[ \, \delta ( {\hat r}^3 - 1 )  \, + \, \delta ( - {\hat r}^3 - 1 ) \, ] \,
\exp ( i \, k \, z ) \,  \cr
\noalign{\noindent and }
{\cal H}_+ \, \exp ( i \, k \, z )
&= \,  + \, i \, \exp ( i \, k \, z ) \qquad \qquad  z > 0   \cr
&\, \; \; \; - \, i \, \exp ( i \, k \, z ) \qquad \qquad   z < 0  \cr
{1 \over \sqrt{r} } \, {\cal H}_+   \sqrt{r} \,  w_{k^3}
&=\, {i \over \sqrt{ 2 \pi k r } } \,
[ \, \delta ( {\hat r}^3 - 1 )  \, - \, \delta ( - {\hat r}^3 - 1 ) \, ] \,
\exp ( i \, k \, z ) \,   \cr
- \, {1 \over \sqrt{r} } \,
\partial_r \, {\cal H}_+   \sqrt{r} \,  w_{k^3}
&=\, {1 \over \sqrt{ 2 \pi k r } } \, k \, {\hat r}^3 \,
[ \, \delta ( {\hat r}^3 - 1 )  \, - \, \delta ( - {\hat r}^3 - 1 ) \, ] \,
\exp ( i \, k \, z ) \,   \cr
&=\, {1 \over \sqrt{ 2 \pi k r } } \, k \,
[ \, \delta ( {\hat r}^3 - 1 )  \, + \, \delta ( - {\hat r}^3 - 1 ) \, ] \,
\exp ( i \, k \, z ) \, = \, k \, w_{k^3}  \, . \cr
      }  $$
The operator components $ \partial_r \, , \, {\cal H}_+ \, $
of $ \overline p^0 \, $ do not commute.
If in the working out of (22) we take $ k \, $ to the right instead of to the
left, we arrive at
$$ \eqalignno{
\overline p^0 \, = \, \tilde{\cal U} \,  k \, {\cal U} \,
&= \, - \, {1 \over 2 \sqrt{r} } \,
\left[ \, ( {\cal H}_e + {\cal H}_o )  \,
- \,   ( {\cal H}_e - {\cal H}_o )  \, {\cal P} \, \right]  \, \partial_r \,
\sqrt{r} \, \cr
&= \, - \, {1 \over \sqrt{r} } \, {\cal H}_-  \partial_r \,  \sqrt{r} \,
      & (24) \cr
   }  $$
where
$$ \eqalignno{
{\cal H}_-  f({\bf r})
&\equiv \, {1 \over 2} \, \left[ \, ( {\cal H}_e + {\cal H}_o )  \,
- \,  ( {\cal H}_e - {\cal H}_o )  \, {\cal P} \, \right] \, f({\bf r})
  &  (25a) \cr
&= \,  { 1 \over \pi } \, \int_{- \infty}^\infty
 \, \hbox{sgn} (\lambda ) \,  {  f( \lambda {\bf r}) \over \lambda - 1 }  \;
d\lambda \, . &  (25b)   \cr
     }  $$
As
$$ {\cal H}_- {\cal H}_+ = {\cal H}_+ {\cal H}_- = \, - \, 1 \, ,$$
which can be verified their definitions (23a), (25a) together with
$ {\cal H}_e {\cal H}_o = {\cal H}_o {\cal H}_e = \, - \, 1 \, ,$ then
$$ {\overline p^0}^2
= \, {1 \over \sqrt{r} } \, \partial_r \, {\cal H}_+
{\cal H}_-  \partial_r \,  \sqrt{r} \,
= \, - {1 \over \sqrt{r} } \, \partial_r^2 \,  \sqrt{r}
= {\bf \overline p}^2 \, $$
as necessary.
Also with the adjoint relations
$  ( {1 \over \sqrt{r} } \, {\cal H}_{ e \atop o } \, \sqrt{r} )^{\dag } \,
= \, - \, {1 \over \sqrt{r} } \, {\cal H}_{ o \atop e } \, \sqrt{r} \, $
then from (23a), (25a)
$$  ( {1 \over \sqrt{r} } \, {\cal H}_{\pm} \, \sqrt{r} )^{\dag } \,
= \, - \, {1 \over \sqrt{r} } \, {\cal H}_{\mp} \, \sqrt{r} \, . $$

\vskip 0.1 in

\noindent {\bf 2.  Wave equations } \hfil\break
We can write a wave equation substituting
$ {\bf \overline p} \, $ for the usual $ {\bf p} \, :$
$$\eqalignno{
&[ ( i \partial_t )^2 \, - \, {\bf \overline p}^2 ] \psi = 0 \, \cr
&[ - \partial_t^2  + r^{-1/2} \partial_r^2 \, r^{1/2}  \,  ] \psi = 0 \,
           &  (26) \cr
  }  $$
which is essentially the wave equation
$ - \square \psi \equiv [ - \partial_t^2  + {1 \over r} \partial_r^2 r
- {1 \over r^2} \, L^2  ] \psi = 0 \, $ with the
operator component $ {1 \over r^2} \, L^2 \, $ excluded.
This modified wave equation (26) has solutions
$ \psi_{\bf k} , \psi^*_{\bf k} \, $ which are eigenfunctions of
$ {\bf \overline p} \, :$
$$ \psi_{\bf k} = \exp ( - i k t  ) \, w_{\bf k} \, ,
\qquad \qquad \psi^*_{\bf k} = \exp ( i k t  ) \, w^*_{\bf k} \, .
\eqno (27) $$
Both the states $ \psi_{\bf k} , \psi^*_{\bf k} \, $
are waves localized along the propagation axis
$ {\bf r} =  \lambda \, {\bf \hat{k}} \;
( - \infty < \lambda < \infty ) \, $ and proceeding in the
$ + {\bf \hat{k}} \, $ direction, and accords with our everyday experience
of photons being localized along their propagation axis.
If $ \psi \, $ satisfies (26) then there is a conserved density
$$ \eqalignno{
\sigma &= i \, ( \psi^* \, \partial_t \psi
- ( \partial_t \psi^* ) \, \psi  \cr
  }  $$
However this density is indefinite.

We will from now consider the first order evolution equation
$$ i  \partial_t \psi = \overline p^0 \psi
\equiv \, - \, r^{-1/2} \partial_r \, {\cal H}_+ \, r^{1/2} \,
\psi \, . \eqno (28) $$
Now only the positive energy eigenstates $ \psi_{\bf k} \, $ of (27)
are solutions and the negative energy components
$ \psi^*_{\bf k} \, $ are excluded.
Consider the non-negative density $ \rho \, $
$$ \rho =  {1 \over r} \, [ \psi^* \, \psi \,
+ \, ( r^{-1/2} {\cal H}_+ \, r^{1/2} \, \psi^* ) \;
( r^{-1/2} {\cal H}_+ \, r^{1/2} \, \psi ) \, ] \,   $$
then if $ \psi \, $ is a solution of (28)
$$ \eqalignno{
\partial_t \rho
&=  {1 \over r} \, \partial_t \psi^* \, \psi \,
+ \, {1 \over r} \, ( r^{-1/2} {\cal H}_+ \, r^{1/2} \, \psi^* ) \;
( r^{-1/2} {\cal H}_+ \, r^{1/2} \, \partial_t \psi ) + \, \hbox{ CC }   \cr
&= - \, {i \over r} \, ( r^{-1/2} \partial_r {\cal H}_+ \, r^{1/2} \, \psi^* ) \,
\psi \,
+ \, {i \over r} \, ( r^{-1/2} {\cal H}_+ \, r^{1/2} \, \psi^* ) \;
( r^{-1/2} {\cal H}_+ \, {\cal H}_- \, \partial_r r^{1/2} \, \psi )
+ \, \hbox{ CC }    \cr
&= - \, {i } \, ( r^{-1} \partial_r {\cal H}_+ \, r^{1/2} \, \psi^* ) \,
( r^{-1/2} \, \psi ) \,
- \, {i } \, ( r^{-1} {\cal H}_+ \, r^{1/2} \, \psi^* ) \;
( r^{-1} \partial_r r^{1/2} \, \psi )   + \, \hbox{ CC }     \cr
&= {\bx \nabla} \cdot  {\bf \hat{r}} \,
\Big[ - \, i \, ( r^{-1} \, {\cal H}_+ \, r^{1/2} \, \psi^* ) \,
( r^{-1/2} \, \psi ) \,  + \, \hbox{ CC } \, \Big]
\equiv \, - \, {\bx \nabla} \cdot {\bf J} \,  \cr
  }  $$
where CC stands for the complex conjugate terms
and we have used the operator identity
$$  ( r^{-1} \, \partial_r \, r \, A ) \, B \, + \,
A \, ( r^{-1} \, \partial_r \, r \, B )
= {\bx \nabla} \cdot  [ {\bf \hat{r}} \, A \, B \, ] \, . $$
Substituting $ \psi = w_{\bf k} \, $
into the above expressions for the density and current,
both the density and current are evenly distributed along the
$ {\bf k} \, $ axis $ {\bf r} =  \lambda \, {\bf \hat{k}} \, \;
( - \infty < \lambda < \infty ) \, ,$ and the current is uniformly in the
$ \, + \, {\bf \hat{k}} \, $ direction.

\vskip 0.1 in

\noindent {\bf 3. Lorentz transformations } \hfil\break
The momentum space transformation is straightforward.
The $ k^\lambda = ( k  , {\bf k} ) = ( k  , k^1 , k^2 , k^3 ) \, $
transforms in the usual way as a 4-vector, that is under a finite boost
of velocity $ v $ in the $ z $ direction then
$$ \eqalign{
k'^1 &= k^1 \, , \qquad k'^2 =  k^2 \, ,
\qquad  k'^3 =  \gamma k^3 - \gamma v k \, , \qquad
k' = \gamma k - \gamma v k^3 \, , \qquad  \cr
  } \eqno(29)  $$
and
$$ \eqalign{
\hat k'^1 = \hat k^1 / \gamma ( 1 - v \hat k^3 ) \, , \qquad
\hat k'^2 =  \hat k^2 / \gamma ( 1 - v \hat k^3 ) \, ,
\qquad  \hat k'^3 =  ( \hat k^3 - v ) / ( 1 - v \hat k^3 ) \,   \cr
  } \eqno(30)  $$
with $ \gamma = ( 1 - v^2 )^{- 1/2 } \, .$
The last of (30) is the well known aberration formula,
as putting $ \hat k^3 = - \cos \theta \, $ (light observed from the
direction $ {\bf \hat{k}} \, $ has momentum $ \propto - {\bf \hat{k}} \, )$
then
$$ \cos \theta' =  ( \cos \theta + v ) / ( 1 + v \cos \theta ) \, .$$
The basis state
$  \psi_{\bf k}  = \exp ( - i k t  ) \, w_{\bf k} \, $
transforms as a scalar in momentum space, i.e.
$$ ( \psi_{\bf k} )'
= ( \psi_{\bf k} )_{ {\bf k} \rightarrow {\bf k'}  }
=  \, {1 \over \sqrt{ 2 \pi k' r } } \,
[ \delta ( - {\bf \hat{r}} , {\bf \hat{k'}} ) \,
+ \, \delta ( {\bf \hat{r}} , {\bf \hat{k'}} ) \, ] \,
\exp ( - i k' t + i {\bf k'} \cdot {\bf r} ) \,  \eqno (31) $$
where $ {\bf k'} \, $ is given by (29).
From (31) can be inferred the frequency change from the exponential term,
and also the change of the propagation direction
$ {\bf \hat k} \, $  which agrees with the aberration formula.

The $ {\bf k} \, $ transformation is generated by the boost and rotation
operators in momentum space which are
$$ {\bf N}_k =  i \, k \, (\partial / \partial {\bf k} )  \, ,
\qquad \qquad {\bf L}_k = - \, i \, ( {\bf k} \times \,
\partial / \partial {\bf k} )  \, \eqno (32) $$
then (29) is $ {\bf k'} = \exp ( v \, N^3 ) \, {\bf k} \, .$
In configuration space the boost transformation will be complicated due to the
fact that $ \overline p^0 \, $ is a non-local operator, so there is no simple
transformation of the configuration space $ {\bf r} \, $ coordinates
corresponding to (29). The Lorentz boost generator
$ {\bf N}  \, $
must satisfy
$$ [ {\bf N} \, , \, \overline p^0  \, ] =   i \, {\bf \overline p} \,
\qquad \qquad [ N^a \, , \, \overline p^b \, ]
=  i \, \delta^{a b } \, \overline p^0  \,
\eqno (33) $$
which implies that $ {\bf N}  \, $ as well as $ \overline p^0 \, $ is
a non-local (integral) operator.
The operators generating the Lorentz transformations in configuration
space are
$$ \eqalign{
{\bf N}  &= ( r^{-1/2} {\cal H}_- \, r^{1/2} ) \,
( r \, {\bx \nabla} \, - \, 2 \, {\bf \hat r} \, \partial_r r ) \,
= ( r \, {\bx \nabla} \, - \, 2 \, {\bf \hat r} \, \partial_r r ) \,
( r^{-1/2} {\cal H}_+ \, r^{1/2} ) \,  \cr
{\bf L}  &= - \, i \, {\bf r} \, \times {\bx \nabla} \,  \cr
       }   \eqno (34)  $$
which are Hermitian with respect to the $ 1 / r \, $ inner product space (9).
That the $ {\bf N} \, $ defined above satisfy the relations (33), and that
the Lorentz group relation $ [ N^a , N^b ] = - i \epsilon^{abc} L^c \, $
is also satisfied we leave to the appendix.
An infinitessimal
boost $ \delta {\bf v} \, $ causes a change in the wavefunction
$ \psi \rightarrow ( 1 + \, i \, \delta {\bf v} \cdot {\bf N} ) \, \psi \, .$
In the appendix we calculate
$ {\bf N} \, w_{\bf k} \, $ and show that the combined transformation
$ {\bf N} \, w_{\bf k} + {\bf N}_k \, w_{\bf k} = 0 \, ,$
which means that $ w_{\bf k} \, $ is a Lorentz scalar.

This completes our investigation of the spin-zero operators
$ \overline p^0 \, , {\bf \overline p} \, , {\bf N} \, , {\bf L} \, $
which together obey the Poincar\'e group commutation relations.

\vskip 0.1 in

\noindent {\bf 4. Spin-${1 \over 2}$ and spin-1 representations }
\hfil\break
Massless spin-${1 \over 2}$ and spin-1 Hamiltonians can be constructed
which are differential operators, in contrast to the non-local
$ \overline p^0 \, $ Hamiltonian for the spin-zero case.
The spin-${1 \over 2}$ massless
(neutrino) Hamiltonian equation
$$ i \, \partial_t \Psi = {\bmit \sigma } \cdot {\bf \overline p} \, \Psi
=  - \, i \, ( {\bmit \sigma } \cdot {\bf \hat{r}} ) \,
( \partial_r + {1 \over 2 r } ) \, \Psi  \,  \eqno (35) $$
has as eigensolutions either column of the $ 2 \times 2 \, $ matrix
$ e^{ \pm i k t } ( {\bmit \sigma } \cdot {\bf k} \, ) \,
w_{\bf k} \, .$
Formally one can also construct a spin-${1 \over 2}$ Dirac-like equation
with mass
$ \, i \, \partial_t \Psi = ( {\bmit \alpha } \cdot {\bf \overline p} \,
+ \beta \, m ) \, \Psi \; ,$
which has as eigensolutions a column of the $ 4 \times 4 \, $ matrix
$ \, [ \exp ( - i ( m^2 + k^2 )^{1/2} \, t ) ( {\bmit \alpha } \cdot {\bf k} \,
+ \beta \, m ) \, w_{\bf k}  ] \, ,$
however in the rest state when $ {\bf k} = 0 \, $ then $ w_{\bf k} \, $
is undefined.

The source-free electromagnetic field is a massless spin-1 field.
Defining the complex field
$ {\bf F} \equiv {\bf E} + i {\bf B} \, ,$ then the source-free
Maxwell's equations are
$$ [ \partial_t + i {\bx \nabla} \, \times \, ] \,
{\bf F}  = 0 \, ,\quad \qquad
{\bx \nabla} \cdot  {\bf F}  = 0 \, . $$
which we will replace by
$$ [ \partial_t -  {\bf \overline p} \, \times \, ] \,
{\bf F}  = 0 \, ,\quad \qquad
{\bf \overline p}  \cdot  {\bf F}  = 0 \, . \eqno (36)  $$
A solution of (36) is
$$  {\bf F}  = e^{ - i k t } \, ( w_{k^3} , i \, w_{k^3} , 0 )
 $$
which is a wave localized along the $ z \, $ axis
proceeding in the $ + z \, $ direction.

\vskip 0.1 in

\noindent {\bf 5. Outlook } \hfil\break
The attraction of the $ {\overline p}^\lambda \, $ operators is that their
eigenfunctions conform with our experience of photons:
that they appear to be localized along their propagation axis.
On the other hand the chosen origin of
coordinates plays a more significant role in this formalism -
indeed a shifted origin (or observer) may not register
the localized wave at all. In the usual case the exponentiated momentum
operator $ \{ \exp ( {\bf a} \cdot {\bf p} ) \, \} \, $
has a clear geometrical role: in shifting the coordinates by
$ {\bf a} \, .$
By contrast the meaning of the operator
$ \{ \exp ( {\bf a} \cdot {\bf \overline p} ) \, \} \, $ is obscure.

Mathematically the the $ {\overline p}^\lambda \, $ operators are interesting
because they demand continuity through the origin, in contrast to the usual
radial operators whose use is limited apart from describing sources or
sinks at the origin.

\vskip 0.2 in

\noindent {\bf Appendix } \hfil\break
{\noindent \it That the operator $ {\bf \tilde p} \, $ is symmetric with respect
to the inner product } (4)   \hfil\break
Expanding the inner product
$$ \eqalignno{
\Big\langle \phi ( {\bf r} ) \, \big| \,
{\bf \tilde p} \psi ( {\bf r} ) \, \Big\rangle
&\equiv \int_\Omega d\Omega \int_{r \rightarrow 0 }^\infty dr \; r^2 \,
\phi^* ( {\bf r} )  \, [ - \, i \, {\bf \hat{r}}  r^{- 1 } \, \partial_r \, r
\psi ( {\bf r} ) \, ] \,  \cr
&= \int_\Omega d\Omega \int_{r \rightarrow 0 }^\infty dr \; r^2 \,
[ - \, i \, {\bf \hat{r}}  r^{- 1 } \, \partial_r \, r
\phi ( {\bf r} ) \, ]^* \, \psi ( {\bf r} )  \,
- \, i\, \int_\Omega d\Omega \; {\bf \hat{r}}
\Big|_{r \rightarrow 0 }^\infty
\, ( r \phi^* ( {\bf r} )  ) \, \, ( r \, \psi ( {\bf r} )  ) \, \Big| \, \cr
&= \Big\langle {\bf \tilde p} \phi ( {\bf r} ) \, \big| \,
\psi ( {\bf r} ) \, \Big\rangle
- \, i\, \int_\Omega d\Omega \; {\bf \hat{r}}
\Big[ \, ( r \phi^* ( {\bf r} )  ) \, \, ( r \, \psi ( {\bf r} )  ) \,
\Big]_{r \rightarrow 0 } \, & (A1)  \cr
  }  $$
assuming the $ ( r \phi  ) , \, ( r \, \psi  ) \, $ tend to zero sufficiently
fast at infinity. But with the condition (3)
$$ r \, \psi ( \epsilon {\bf r} ) -  r \,\psi ( - \epsilon {\bf r} )
\rightarrow 0 \, \qquad \qquad \hbox{as } \qquad \epsilon \rightarrow 0 \, $$
then the last term on the RHS of (A1) is zero, as the contributions
form opposite sides of the origin cancel. Thus $ {\bf \tilde p} \, $ is symmetric.
Note that it is the presence of the $ {\bf \hat{r}}\, $ within the
$ {\bf \tilde p} \, $ operator that leads to symmetry, by contrast the
operator $ p_r = - \, i \,  r^{- 1 } \, \partial_r \, r \, $ has the more
onerous requirement for symmetry that
$ \, r \, \psi ( r )
\rightarrow 0 \, \quad  \hbox{as } \quad r \rightarrow 0 \, ,$
which is the underlying reason why $ p_r \, $ is not Hermitian
(as discussed by Hellwig p176 [3]).

{\noindent \it The delta function on the unit sphere } \hfil\break
In spherical coordinates
$$ \eqalignno{
{\bf \hat r}  &= ( \sin \theta \cos \phi \, , \,
\sin \theta \sin \phi \, , \, \cos \theta \, ) \,  \cr
{\bf \hat k}  &= ( \sin \theta_k \cos \phi_k \, , \,
\sin \theta_k \sin \phi_k \, , \, \cos \theta_k \, ) \,  \cr
\noalign{\noindent then }
\delta ( {\bf \hat r} \, , \, {\bf \hat k} \, )
&\equiv \delta ( \cos \theta \, - \, \cos \theta_k \, ) \;
\delta ( \phi - \phi_k \, )  \,              & (A2) \cr
\delta ( - {\bf \hat r} \, , \, {\bf \hat k} \, )
&= \delta ( \cos \theta  \, + \, \cos \theta_k \, ) \;
\delta ( \phi + \pi - \phi_k \, )  \, .            & (A3) \cr
       }   $$

{\noindent \it Orthogonality and completeness of the $ u_{\bf k} \, $
\hfil\break }
Expanding out (7a)
$$ \eqalignno{
\int d^3 {\bf r} \; u^*_{\bf k} ( {\bf r} ) \, u_{\bf k'}  ( {\bf r} )
&= \, \int_\Omega d\Omega \int_{0}^\infty dr \;
 {1 \over 2 \pi \,k k' } \, [ \delta ( - {\bf \hat{r}} ,
{\bf \hat{k}} ) \, \exp ( i k r ) \, + \, \delta ( {\bf \hat{r}} ,
{\bf \hat{k}} ) \, \exp ( - i k r ) ] \, \times \cr
& \qquad \qquad  \qquad \qquad [ \delta ( - {\bf \hat{r}} ,
{\bf \hat{k'}} ) \, \exp ( - i k' r ) \, + \, \delta ( {\bf \hat{r}} ,
{\bf \hat{k'}} ) \, \exp ( i k' r ) ] \,    \cr
&= \,  {1 \over 2 \,k k' } \, \int_\Omega d\Omega \;
[ \delta ( - {\bf \hat{r}} , {\bf \hat{k}} ) \,
\delta ( - {\bf \hat{r}} , {\bf \hat{k'}} ) \,
+ \, \delta ({\bf \hat{r}} , {\bf \hat{k}} ) \,
\delta ({\bf \hat{r}} , {\bf \hat{k}} ) ] \;  \delta ( k - k' ) \,  \cr
&= \,  {1 \over k k' } \;  \delta ({\bf \hat{k}} , {\bf \hat{k'}} ) \,
\; \delta ( k - k' ) \, \equiv \, \delta ({\bf k} - {\bf k'} ) \, . & (A4)  \cr
  }  $$

{\noindent \it The boost operator {\bf N} of (34) } \hfil\break
\noindent First consider the operator
$$ \eqalignno{
{\bf N'} &\equiv  - \, i ( r \, {\bx \nabla} \,
- \, 2 \, {\bf \hat{r}} \, \partial_r r ) & (A5) \cr
   } $$
and defining
$$ \eqalignno{
s^\lambda &= ( - \, {1\over r} \, , \, {{\bf \hat{r}} \over r} \, )  & (A6) \cr
  }  $$
the following commutation relations may be directly verified
$$ [ {\bf N'} \, , \, s^0 \, ] =   i \, {\bf s} \, ,
\quad \qquad [ N'^a \, , \, s^b \, ] =  i \, \delta^{a b } \, s^0 \, .
\eqno (A7) $$
Also the operator $ - i \partial_r r \, $ commutes with $ {\bf N'} \, $
so that
$$ \eqalignno{
t^\lambda &= ( i {1\over \sqrt{r}} \partial_r \sqrt{r} \, , \,
- i \, {\bf \hat{r}} \, {1\over \sqrt{r}} \partial_r \sqrt{r} \, )
= ( i {1\over \sqrt{r}} \partial_r \sqrt{r} \, , \,
{\bf \overline p} \, )   & (A8) \cr
  }  $$
also obeys (A7) with $ t^\lambda \, $ instead of $ s^\lambda \, .$
Finally we multiply $ t^0 \, $ on the right by
$ ( i { 1 \over \sqrt{r}} {\cal H}_+ \sqrt{r} \, ) \, $ and
$ {\bf N'} \, $ on the left by
$ ( i { 1 \over \sqrt{r}} {\cal H}_- \sqrt{r} \, ) \, $
to make the new operators $ {\overline p}^0 \, , \; {\bf N} \, ,$
which recalling $ {\cal H}_+ {\cal H}_- = - 1 \, $
also satisfy the relations corresponding to (A7), i.e.
$$ \eqalign{
{\bf N}  &= ( {1\over \sqrt{r}} {\cal H}_- \sqrt{r} ) \,
( r \, {\bx \nabla} \, - \, 2 \, {\bf \hat r} \, \partial_r r ) \,
= ( r \, {\bx \nabla} \, - \, 2 \, {\bf \hat r} \, \partial_r r ) \,
( {1\over \sqrt{r}} {\cal H}_+ \sqrt{r} ) \,  \cr
{\overline p}^0
&= - \, {1\over \sqrt{r}} \, \partial_r \, {\cal H}_+ \sqrt{r} \,
= - \, {1\over \sqrt{r}} \, {\cal H}_- \partial_r \, \sqrt{r} \, , \qquad \quad
{\bf \overline p} = - i \, {\bf \hat{r}} \,
{1\over \sqrt{r}} \partial_r \, \sqrt{r} \,  . \cr
       }   \eqno (A9)  $$
Finally the fact that
$$ [ N^a \, , \,  N^b \, ] =  - i \, \epsilon^{a b c} \, L^c \, \eqno (A10) $$
follows from
$ [ N'^a \, , \,  N'^b \, ] =  - i \, \epsilon^{a b c} \, L^c \, $
where $ {\bf L}  = - \, i \, {\bf r} \, \times {\bx \nabla} \, .$

{\noindent \it To show that $ {\bf N} \, w_{\bf k} \,
+ \, {\bf N}_k \, w_{\bf k} \, = 0 \, $ } \hfil\break
\noindent We will consider the boost operators $ N^3 \, , N_k^3 \, $
of (32), (34), and
recalling $ w_{\bf k} \, $ from (10) we have in spherical coordinates
$$ \eqalignno{
N_k^3 &= \, i \, \cos \theta_k \, k \, \partial_k
    - \, i \, \sin \theta_k \, \partial_{\theta_k} \; \cr
N^3 &= ( - \,  \cos \theta \, r \, \partial_r
    - \, \sin \theta \, \partial_{\theta} )
\, ( {1\over \sqrt{r}} {\cal H}_+ \sqrt{r} \, )  \;  \cr
w_{\bf k} &=\, {1 \over 2 \sqrt{ \pi k r } } \,
\big[ \delta ( \cos \theta \, - \, \cos \theta_k \, ) \;
\delta ( \phi - \phi_k \, )  \,   ) \, \exp ( i k r ) \, \cr
&\qquad \qquad \qquad + \, \delta ( \cos \theta \, + \, \cos \theta_k \, ) \;
\delta ( \phi + \pi - \phi_k \, ) \, \exp ( - i k r ) \, \big] \,    \cr
( {1\over \sqrt{r}} {\cal H}_+ \sqrt{r} \, )  w_{\bf k}
&=\, {i \over 2 \sqrt{ \pi k r } } \,
\big[ \delta ( \cos \theta \, - \, \cos \theta_k \, ) \;
\delta ( \phi - \phi_k \, )  \,   ) \, \exp ( i k r ) \, \cr
&\qquad \qquad \qquad - \, \delta ( \cos \theta \, + \, \cos \theta_k \, ) \;
\delta ( \phi + \pi - \phi_k \, ) \, \exp ( - i k r ) \, \big] \, .   \cr
   }   $$
Then
$$ \eqalign{
N_k^3 \, w_{\bf k}
=\, &+ \, i \, \cos \theta_k \,  \delta ( \cos \theta \, - \, \cos \theta_k \, ) \;
\delta ( \phi - \phi_k \, )  \, k \, \partial_k \;
[ {\exp ( i k r )  \over 2 \sqrt{ \pi k r } } \, ] \, \cr
&- \, i \;   {\exp ( i k r )  \over 2 \sqrt{ \pi k r } } \,
\delta ( \phi - \phi_k \, )  \,   ) \,
\sin \theta_k \; \partial_{\theta_k} \,
[ \delta ( \cos \theta \, - \, \cos \theta_k \, )  ] \; \cr
&+ \, i \, \cos \theta_k \, \delta ( \cos \theta \, + \, \cos \theta_k \, ) \;
\delta ( \phi + \pi - \phi_k \, ) \, k \, \partial_k \;
[ {\exp ( - i k r ) \over 2 \sqrt{ \pi k r } } \, ] \, \cr
&- \, i \; {\exp ( - i k r )  \over 2 \sqrt{ \pi k r } } \,
\delta ( \phi + \pi - \phi_k \, )  \,   ) \,
\sin \theta_k \; \partial_{\theta_k} \,
[ \delta ( \cos \theta \, + \, \cos \theta_k \, )  ] \;     \cr
   }  \eqno   (A11)   $$
and
$$ \eqalign{
N^3 \, w_{\bf k}
=\, &- \, i \, \cos \theta \,  \delta ( \cos \theta \, - \, \cos \theta_k \, ) \;
\delta ( \phi - \phi_k \, )  \, r \, \partial_r \;
[ {\exp ( i k r )  \over 2 \sqrt{ \pi k r } } \, ] \, \cr
&- \, i \;  {\exp ( i k r )  \over 2 \sqrt{ \pi k r } } \,
\delta ( \phi - \phi_k \, )  \,   ) \,
\sin \theta \, \partial_{\theta} \,
[ \delta ( \cos \theta \, - \, \cos \theta_k \, )  ] \; \cr
&+ \, i \, \cos \theta \, \delta ( \cos \theta \, + \, \cos \theta_k \, ) \;
\delta ( \phi + \pi - \phi_k \, ) \, r \, \partial_r \;
[ {\exp ( - i k r ) \over 2 \sqrt{ \pi k r } } \, ] \, \cr
&+ \, i \; {\exp ( - i k r )  \over 2 \sqrt{ \pi k r } } \,
\delta ( \phi + \pi - \phi_k \, )  \,   ) \,
\sin \theta \, \partial_{\theta} \,
[ \delta ( \cos \theta \, + \, \cos \theta_k \, )  ] \, .     \cr
   }  \eqno   (A12)   $$
Now if we add (A12) and (A12) together then the components cancel in order.

\beginsection References

\frenchspacing
\noindent
\item{[1]} Liboff R. L. et al, ``On the radial momentum operator,'' Am. J. Phys.
{\bf 41}, 976-980 (1973)
\item{[2]} Rooney P. G., ``On the ranges of certain fractional integrals,''
Can. J. Math. {\bf 24}, 1198-1216 (1972)
\item{[3]} Hellwig G.,  {\it Differential operators of mathematical physics}
(Addison-Wesley, Reading, 1964)

\bye

\end